\documentclass{aastex}          
\usepackage{spr-astr-addons}    
\usepackage{url}
\usepackage{verbatim}

\RequirePackage{color}

\begin{document}
%
\title{The thermodynamics of a gravitating vacuum}

\shorttitle{<Short article title>}
\shortauthors{<Autors et al.>}

\author{M. Heyl\altaffilmark{}}
\affil{Deutsches Zentrum f{\"u}r Luft und Raumfahrt (DLR), K{\"o}nigswinterer Str. 522 - 524, 53227 Bonn, Germany}
\and 
\author{H.J. Fahr\altaffilmark{}}
\author{M. Siewert\altaffilmark{}}
\affil{Argelander Institut f{\"u}r Astronomie, Universit{\"a}t Bonn, Auf dem H{\"u}gel 71, 53121 Bonn, Germany}


\begin{abstract}
In the present days of modern cosmology it is assumed that the main ingredient to cosmic energy
presently is vacuum energy with an energy density $\epsilon_\mathrm{vac}$ that is constant over the
cosmic evolution. In this paper here we show, however, that this assumption
of constant vacuum energy density is unphysical, since it conflicts with the
requirements of cosmic thermodynamics. We start from the total vacuum energy
including the negatively valued gravitational binding energy and show
that cosmic thermodynamics then requires that the cosmic vacuum energy density can only vary with cosmic scale $R=R(t)$
according to $\epsilon _\mathrm{vac}\sim R^{-\nu }$ with only two values of $\nu$
being allowed, namely $\nu _\mathrm{1}=2$ and $\nu _\mathrm{2}=5/2$. We then discuss
these two remaining solutions and find, when requiring a universe with
a constant total energy, that the only allowed power index
is $\nu _\mathrm{1}=2$. We discuss the consequences of this scaling of $\epsilon
_\mathrm{vac}$ and show the results for a cosmic scale evolution of a
quasi-empty universe like the one that we are presently faced by.
\end{abstract}

\keywords{Cosmic vacuum energy density -- Friedmann equations -- Thermodynamics}

\section{Introduction}
We start this paper asking why at all should a vacuum gravitate or influence
spacetime geometry? This question is perhaps worth to be asked, since, if
vacuum, expressis verbis, represents 'nothing' in a physical sense, then
it should not do anything, especially should not gravitate, unless it is
wrongly defined. Modern physics nowadays argues, however, that a vacuum
cannot be energy-less, but is loaded with energy, or, due to the energy-mass
equivalence, is mass-loaded. Masses, on the other hand, do in general
gravitate, unless something else compensates for that. But how could sources
of gravity be compensated, unless perhaps by anti-masses which are not known
to exist?

The General Relativistic action of a vacuum in general is taken into account
by a fluid-like hydrodynamical energy-momentum tensor $T_{\mu \nu }^{vac}$
which describes how the vacuum, due to its pressure $p_{vac}$ and its mass
energy density $\rho _{vac},$ acts as source of spacetime geometry (see e.g.
Goenner 1996). If in addition vacuum energy density $\epsilon_{vac}=\rho
_{vac}c^{2}$ is assumed to be constant, as done in present-day standard
cosmologies (see Perlmutter et al. 1999; Bennett \& Halpern 2003), then this
induces the relation $p_{vac}=-\epsilon _{vac}$ (see e.g Peebles \& Ratra
2003) and leads to the following geometrical source tensor (see e.g.
Overduin \& Fahr 2003) $T_{\mu \nu }^{vac}=\rho _{vac}c^{2}g_{\mu \nu }$,
where $g_{\mu \nu }$denotes the metric tensor.

This term $T_{\mu \nu }^{vac}$, since being isomorphal, can be taken
together with the term due to Einstein's cosmological constant $\Lambda _{0}$
(Einstein 1917). If both terms are placed on the right-hand side of the
GRT\ field equations, while Einstein placed his term on the left hand side,
they can be put together representing an 'effective' cosmological constant 
$\Lambda_{eff}$ given by (Overduin \& Fahr 2001; Fahr 2004)

\begin{equation}
\Lambda_{eff}=\frac{8\pi G}{c^{2}}\rho _{vac,0}-\Lambda_{0}.  \label{2}
\end{equation}

Now one can draw the following conclusion: A completely empty, matter-free
space, not doing anything in terms of gravity, is realized, if ,evident from
the above, $\Lambda_{eff}$ just vanishes, i.e. the cosmological term $%
\Lambda _{0}$ just compensates the vacuum energy density of empty
space whatever maybe its value (e.g. see Zeldovich 1968; Carroll, Press \& Turner 1992).

Interestingly, very similar ideas have come up in papers by Sola (see
Sola, 2013, 2014) who expresses the fact that in order to settle down the
spacetime geometry of a pure vacuum to a nongravitating Minkowskian
spacetime within a covariant general-relativistic field theory the effective
vacuum energy of this empty space has to vanish.

In the presence of real matter the argumentation, however, is much more
complicate as we have discussed at several places in the literature
(Overduin \& Fahr 2001; Fahr 2004; Fahr \& Heyl 2007a, 2007b; Fahr \&
Sokaliwska 2012). Especially it is then highly questionable whether under
such conditions a constant vacuum energy density can at all be expected as
an option.

If under these perspectives it could be assumed, that only the energy
difference between the matter-polarized and the empty vacuum gravitates then
some interesting new conclusions could be drawn. It then means that in a
matter-filled universe the effective quantity representing the action of the
vacuum energy density is given by:

\begin{equation}
\Lambda_{eff}=\frac{8\pi G}{c^{2}}(\rho _{vac}-\rho _{vac,0}).  \label{3}
\end{equation}

The above formulation expresses that in a matter-filled universe only the
difference between the values of the vacuum energy densities $\rho _{vac,0}$
of empty space and $\rho _{vac}$ of matter-polarized space gravitates, i.e.
the spacetime geometry only reacts to the difference of these vacuum
energies.

Even under these new prerequisites it is nevertheless not the most natural
assumption, that vacuum energy density $\epsilon _{vac}=\rho _{vac}c^{2}$
should be considered as a time-independent quantity. This is because the
unit of volume is not a cosmologically relevant quantity, and vacuum energy
density neither is. It would probably appear more reasonable to assume that
the energy load of any homologously comoving proper volume does not change
with cosmic expansion, i.e. that rather just this proper-energy is constant.
This demand, however, means that the true constant quantity, instead of the
vacuum energy density $\epsilon _{vac}$, is

\begin{equation}
e_{vac}=\epsilon _{vac}\sqrt{-g_{3}} d^{3}V  \label{4}
\end{equation}

where $g_{3}$ is the determinant of the 3d-space metric which in case of a
Robertson-Walker geometry is given by

\begin{equation}
g_{3}=g_{11}g_{22}g_{33}=-\frac{1}{(1-Kr^{2})}R^{6}r^{4}\sin ^{2}\vartheta
\label{5}
\end{equation}

with $K$ denoting the curvature parameter, the $R=R(t)$ determining the
time-dependent scale of the universe, and the differential 3-space volume
element in normalized polar coordinates given by

\begin{equation}
{d}^{3}V={d}r {d}\vartheta  {d}\varphi. \label{6}
\end{equation}

This then leads to the following request

\begin{eqnarray}  
e_{vac} =\epsilon _{vac}\sqrt{R^{6}r^{4}\sin ^{2}\vartheta /(1-Kr^{2})}%
{d}r{d}\vartheta {d}\varphi =\cr
\epsilon _{vac}\frac{R^{3}}{\sqrt{1-Kr^{2}}}r^{2}\sin \vartheta
{d}r{d}\vartheta {d}\varphi ={const.}  \label{7}
\end{eqnarray}

which\ evidently leads to a variability of the vacuum energy density $%
\epsilon _{vac}$in the form

\begin{equation}
\epsilon _{vac}=\rho _{vac}c^{2}\sim R(t)^{-3}.  \label{8}
\end{equation}

In the following paper we shall now throw some new light on the variability
of $\epsilon _{vac}$ that must be expected. We therefore study the behavior
of the vacuum energy density $\epsilon _{vac}$ with the scale $R(t)$ of the
universe from a thermodynamical view.

\section{Thermodynamics of the cosmic vacuum}
In the following cosmological considerations we treat the cosmic vacuum by
quantities denoting its vacuum energy density $\varepsilon _{vac}$ and its
associated vacuum pressure\ $p_{vac}$, like done in case of \ a
hydrodynamic fluid which in general relativity theory is described by the
following fluid-type hydrodynamical energy-momentum tensor (see e.g
Goenner 1996; Overduin \& Fahr 2001; Blome, Hoell \& Priester 2002; Fahr 2004)

\begin{equation}
T_{\mu \nu }^{vac}=(\rho _{vac}c^{2}+p_{vac})U_{\mu }U_{\nu }-p_{vac}\ast
g_{\mu \nu }  \label{9}
\end{equation}

where $\varepsilon _{vac}=\rho _{vac}c^{2}$ and $p_{vac}$ are energy density
and pressure of the vacuum, $U_{i}$ denote the components of the fluid
four-velocity, and $g_{\mu \nu }$ is the four-space metric tensor.

In order to use the above energy-momentum tensor in the frame of the general
relativistic field equations one needs to know, how $\rho _{vac}$ and $%
p_{vac}$ are related to each other and how they are dependent on spacetime
coordinates. For that purpose we want to use the well known thermodynamic
equation that relates the internal volume energy with the work expended at
the expansion of that volume. In its easiest form for a Robertson-Walker
symmetric universe with curvature $K=0$ this equation for a sphere of scale 
$R=R(t)$ is given by (see Goenner 1996):

\begin{equation}
\frac{4\pi }{3}\frac{{d}}{dR}(\varepsilon _{vac}R^{3})=-p_{vac}\frac{4\pi }{3}%
\frac{{d}}{{d}R}R^{3}.  \label{10}
\end{equation}

Analogously to a star at its contraction the internal volume energy,
irrelevant whether it is vacuum- or matter-filled, should, however, be
completed by the gravitational self-binding energy, since a vacuum that is
energy-loaded evidently is a source of internal gravity which at all makes
it cosmologically relevant as source of cosmic geometry. If we include the
negatively valued gravitational self-binding energy (see Fahr \& Heyl 2007a, 2007b) into the total internal energy of a cosmic sphere with radius $R$,
then instead of the above relation one obtains the following more complicate
thermodynamic equation:

\begin{eqnarray}
\frac{{d}}{{d}R}[\frac{4\pi }{3}\varepsilon _{vac}R^{3}-\frac{8\pi ^{2}G}{15c^{4}%
}(\varepsilon _{vac}+3p_{vac})^{2}R^{5}]=\cr
-p_{vac}\frac{4\pi }{3}\frac{{d}}{{d}R}%
R^{3}  \label{11}
\end{eqnarray}

which now instead of Eq. (\ref{10}) should define the relation between $%
\varepsilon _{vac}$ and $p_{vac}$ and both their dependences on the scale
parameter $R=R(t)$ which is a function of the cosmic time $t$.

As evident, in this highly symmetric FLRW universe both quantities, i.e.\ $%
\varepsilon _{vac}$ and $p_{vac}$, can only depend on the scale parameter $%
R(t)$. We now try to solve the above equation, following the same way\ as
already used in the case of the more simple, uppermost thermodynamic
Eq. (\ref{10}), namely assuming a power-law dependence of $\varepsilon
_{vac}$ on $R$ in the form $\varepsilon _{vac}\sim R^{-\nu }$ with an
undefined power index $\nu $, and then obtaining for the vacuum pressure the
relation

\begin{equation}
p_{vac}=-\frac{3-\nu }{3}\varepsilon _{vac}.  \label{12}
\end{equation}

Here so far all power indices, especially the cardinal index values $\nu
=0,1,2,3$, were equally allowed, none of them being apriori excluded,
however, the $R$-dependence of $p_{vac}$ and $\varepsilon _{vac}$ turned out
to be identical.

If we now make use of these earlier results (Eq. (\ref{12}), but try to find solutions of
the extended thermodynamic Eq. (\ref{11}) on the basis of these earlier findings
we then obtain:

\begin{eqnarray}
-\frac{4\pi }{3}\frac{3(3-\nu )}{3-\nu }p_{vac}R^{2}=-3\frac{4\pi }{3}%
p_{vac}R^{2}+ \cr
\frac{8\pi ^{2}G}{15c^{4}}\frac{{d}}{{d}R}[(\varepsilon
_{vac}+3p_{vac})^{2}R^{5}]  \label{13}
\end{eqnarray}

which, since the terms left and right of the identity sign cancel, after
replacing $\varepsilon _{vac}$ by $p_{vac}$with Eq. (\ref{12}) leads to the
requirement

\begin{equation}
0=\frac{(6-3\nu )^{2}}{(3-\nu )^{2}}\frac{{d}}{{d}R}(p_{vac}^{2}R^{5}).
\label{14}
\end{equation}

This equation for a completed thermodynamics now evidently is only solved by
two special values of $\nu $, i.e$.$ the requirements:

a: $\nu =\nu _{1}=2$

and

b: $p_{vac}^{2}R^{5}={const}$, i.e.\ by $\nu =\nu _{2}=5/2$

thus now determining, compared to the earlier result, a much more restricted
set of physically possible dependences of $\ p_{vac}$ and $\varepsilon
_{vac} $ on $R.$

\section{Do there exist two competing solutions?}
From the above derivation the two solutions $\nu =\nu _{1}$ and $\nu =\nu
_{2}$ are competing as equally justified, and one could think of taking a
representation of the form

\begin{equation}
\varepsilon _{vac}=\varepsilon _{0,1}(R/R_{0})^{-\nu _{1}}+\varepsilon
_{0,2}(R/R_{0})^{-\nu _{2}}  \label{15}
\end{equation}

as the most general solution. However, without any concrete, specific
physics behind the different forms, how $\varepsilon _{vac}$ reacts to
cosmic scale expansion, this form of a solution is not really satisfying.
Thus we try to restrict the possible power indices even more by looking at
this question from another view.

Requiring a universe where in every instant the positively valued vacuum
energy is compensated by its gravitationally induced self-binding energy, then
, in addition to the above thermodynamic requirement, one has to also
fullfill the following relation (see Fahr \& Heyl 2007a, 2007b) for a
vanishing total vacuum energy

\setlength\mathindent{0mm}
\begin{equation}
\frac{4\pi }{3}(\varepsilon _{vac}+3p_{vac})R^{3}=\frac{8\pi ^{2}G}{15c^{4}}[%
(\varepsilon _{vac}+3p_{vac})^{2}R^{5}].  \label{16}
\end{equation}
\setlength\mathindent{5mm}
We now solve this quadratic equation with respect to the pressure $p_{vac}$
and get the following two solutions:

\begin{equation}
p_{vac,1}=-\frac{1}{3}\varepsilon _{vac}  \label{17}
\end{equation}

and

\begin{equation}
p_{vac,2}=\frac{1}{3}(\frac{5c^{4}}{2\pi GR^{2}}-\varepsilon _{vac}).
\label{18}
\end{equation}

Insertion of Eq. (\ref{17}) or Eq. (\ref
{18}) into Eq. (\ref{11}) results in both cases in one and the same
differential equation for the energy density $\varepsilon _{vac}$ given by:

\begin{equation}
\frac{{d}\varepsilon_{vac}}{{d}R}R +2\varepsilon_{vac}=0  \label{19}
\end{equation}

which has the unique solution:

\begin{equation}
\varepsilon _{vac}=\varepsilon _{vac,0}\frac{R_{0}^{2}}{R^{2}}\sim p_{vac,1,2}  \label{20}
\end{equation}

with $\varepsilon _{vac,0}$  the vacuum energy density at a scale
parameter $R_{0}$, e.g. at the present cosmic time $t_{0}$. Using $%
\varepsilon _{vac}=\rho _{vac}c^{2}$ we finally get from Eq. (\ref{20}) for
the associated cosmic mass density $\rho _{vac}$ of a pure
vacuum-energy-dominated universe which scales according to $R^{-2}$:

\begin{equation}
\rho_{vac}=\rho_{vac,0}\frac{R_0^2}{R^2}.  \label{21}
\end{equation}

Similar results, however derived independently from very different
theoretical reasons, have already been published by Basilakos (2009), Sol\`{a}
(2013), Basilakos et al. (2013) and Sol\`{a} (2014). In these papers it has been
discussed that strictly keeping to covariance requirements of the underlying
general relativistic field equations one can allow for a time-dependence of
the inherent cosmic vacuum energy density $\rho _\mathrm{vac}$ and, as a
leading term, one should preferably consider the following time-dependence
of the vacuum energy density $\rho _\mathrm{vac}=\rho _\mathrm{vac,0}+\alpha
\cdot H^{2}(t)$, where $H=H(t)=\dot{R}/R$ denotes the time-dependent Hubble
constant within a Friedman-Lemaitre cosmology. As the above authors
emphasize, this new setting will help solving many oustanding problems in
the present-day cosmology like triggering a smooth transition from an
initial inflationary expansion powered by very strong vacuum energy density
into a present-day smooth inflation at very low vacuum energy densities of
the order of $\rho _\mathrm{vac,0}\simeq 10^{-29}g/cm^{3}$.

A similar attempt to subject the field equations to more general
scale-invariance requirements has led Scholz (2008) on the basis
of a Weylian scalar-tensor theory also to a term which acts equivalent to
vacuum energy density and which is varying with $(1/R^{2})$ exactly like
derived in our above approach. The question may, however, come up here with
concern to the justification of a scale-invariance requirement applied to
the GRT field equations. Nevertheless, there are hints from many sides that a
scale- or time-dependent vacuum energy term $\rho _\mathrm{vac}=\rho _\mathrm{vac}(t)$ seems to make much sense in cosmology.

\section{Friedmann-Lema\^{i}tre equations for a $R^{-2}$-scaling of $\rho
_{vac}$}
The Friedmann equations provide a relationship between the cosmic scale $R$,
its first and second time derivatives $\dot{R}$ and $\ddot{R}$ on one hand,
and the cosmic mass density $\rho $ and its associated pressure $p$ on the
other hand. In the following we investigate a pure vacuum energy filled
universe with curvature $K=0$. The Friedmann equations are then given by:

\begin{equation}
H^{2}(t)=\frac{\dot{R}^{2}}{R^{2}}=\frac{8{\pi }G}{3}\rho _{vac}  \label{22}
\end{equation}

and

\begin{equation}
\frac{\ddot{R}}{R}=-\frac{4{\pi}G}{3c^2}(\rho_{vac} c^2+3p_{vac})  \label{23}
\end{equation}

with $H(t)$ the time dependent Hubble parameter. Insertion of the $R^{-2}$%
-dependent equivalent mass density of the vacuum energy given by Eq. (\ref
{21}) into Eq. (\ref{22}) leads to:

\begin{equation}
H^{2}(t)=\frac{\dot{R}^{2}}{R^{2}}=\frac{8{\pi }G}{3}\rho _{vac,0}\frac{R_0^2%
}{R^2}  \label{33}
\end{equation}

which provides the following result for the expansion velocity $\dot{R}$ of
the scaling factor $R$:

\begin{equation}
\dot{R}=\sqrt{\frac{8{\pi}G \rho_{vac,0}}{3}}R_0 = {const.}  \label{24}
\end{equation}

and thus, if we require $R(t=0)=0$:

\begin{equation}
R=\sqrt{\frac{8{\pi}G \rho_{vac,0}}{3}}R_0 t.  \label{25}
\end{equation}

We now look at the 2. Friedmann equation Eq. (\ref{23}). The calculated
pressure in eq. Eq. (\ref{17}) results in a cosmic acceleration which is
simply zero:

\begin{eqnarray}
\ddot{R}=-\frac{4{\pi}G}{3c^2}(\rho_{vac} c^2+3p_{vac,1})R= \cr
-\frac{4{\pi}G}{3c^2}(\rho_{vac} c^2-3\frac{1}{3}\rho_{vac} c^2)R=0.  \label{26}
\end{eqnarray}

However, the pressure in Eq. (\ref{18}) leads to the following expression:

\begin{eqnarray}
\ddot{R}=-\frac{4{\pi}G}{3c^2}(\rho_{vac} c^2+3\frac{1}{3}\frac{5c^4}{2 \pi
GR^2}- \cr 
3\frac{1}{3}\rho_{vac} c^2)R=-\frac{10c^2}{3R}.  \label{27}
\end{eqnarray}

The result of Eq. (\ref{27}) is in discrepancy with the constant expansion
velocity $\dot{R}$ in Eq. (\ref{24}) which follows from the 1. Friedmann
equation which itself does not depend on the pressure. Thus, since a
constant $\dot{R}$ cannot be realized with Eq. (\ref{27}), we can conclude
that the pressure in Eq. (\ref{18}) and its associated acceleration in Eq. (%
\ref{27}) are of course mathematical solutions of our thermodynamical
equations but not physical ones which are realized in a cosmos with a vacuum
energy density which scales according to $R^{-2}$ and which always leads to $%
\dot{R}={const.}$, i.e. $\ddot{R}=0$. With other words, the correlation
between a vacuum energy density $\epsilon_{vac}\sim R^{-2}$ and its
associated pressure $p_{vac}$ is given by (equation of state):

\begin{equation}
p_{vac}=-\frac{1}{3}\epsilon_{vac}.  \label{34}
\end{equation}

\section{Consequences of the $R^{-2}$-scaling of $\rho _{vac}$ and conclusions}

With the results of the previous chapter for a matter-free, empty universe
dominated by pure vacuum energy and with a curvature parameter $K=0$ (i.e.
a flat vacuum universe) we now look at the Hubble parameter $H(t)$ which is
given for this universe by (see Eqs. (\ref{24}) and (\ref{25}):

\begin{equation}
H(t)=\frac{\dot{R}}{R}=\frac{\sqrt{\frac{8\pi G\rho _{vac,0}}{3}}R_{0}}{\sqrt{\frac{8\pi G\rho _{vac,0}}{3}}R_{0}t}=\frac{1}{t}
\label{28}
\end{equation}

and for the present cosmic time $t_{0}$ leads to $t_{0}=1/H_{0}(t_{0})%
\approx 1,37\cdot 10^{10}{yrs}$  with the presently accepted Hubble parameter $
H_{0}\approx 72{km/s/Mpc}$ (see Bennett et al. 2003).

Furthermore, we can now try to calculate the equivalent of the total,
global vacuum energy content of the universe, i.e. the mass content $M_{vac}$ of such an universe assuming that the extension of the visible universe
is given by the so-called Hubble radius $R_{{H}}$, defined as that cosmic
distance where the cosmic recession velocity $\dot{R}$ equals the velocity
of light $c$ and given by:

\begin{equation}
R_{{H}}=\frac{c}{H(t)}=ct  \label{29}
\end{equation}

with $H(t)$ given by Eq. (\ref{28}). Now, in addition Eq. (\ref{22}) leads us
to the cosmic density:

\begin{equation}
\rho_{vac}=\frac{3H^2}{8 \pi G}=\frac{3}{8 \pi G t^2}  \label{40}
\end{equation}

which is nowadays ($t=t_0$): 
\begin{equation}
\rho_{vac,0}=\frac{3H_0^2}{8 \pi G}=\frac{3}{8 \pi G t_0^2}\approx 10^{-26} 
\frac{{kg}}{{m^3}}.  \label{30}
\end{equation}

Hence we can express the present vacuum mass of the universe by:

\begin{eqnarray}
M_{vac}=\frac{4 \pi}{3}\rho_{vac,0}R_H^3=\frac{4 \pi}{3}\frac{3}{8 \pi G
t_0^2} c^3 t_0^3= \cr
\frac{c^3}{2G}t_0 \approx 10^{53} {kg} \approx 10^{80} m_{p}
\label{31}
\end{eqnarray}

with $m_{p}$ as the mass of the proton. Interestingly, the Eqs. (\ref{30})
and (\ref{31}) show well-known numbers, quite familiar to nowadays
astronomers, namely just numbers for the presently assumed critical mass
density of our universe and the present mass content of the visible
universe, respectively. This may in first glance appear to be completely
casual and be highly astonishing, since with the above we calculated density
and mass of a cosmic vacuum on the basis of a $R^{-2}$-scaling vacuum
energy density, while the numbers that we got are typical for the matter
content of our present universe.

These above results are, however, not judged by the authors of this paper to
be an numerical artifact, but may have the following important reason: We
can take Eq. (\ref{21}) to calculate the equivalent mass density of the
vacuum energy density of the very early universe, i.e. at the Planck time $%
t_{p}$ or the Planck length $R_{H}(t_{p})=r_{p}=ct_{p}$, thereby expressing
the reference scale $R_{0}$ by the present Hubble radius $R_{H,0}=ct_{0}$
(according to Eq. (\ref{29})) and get:

\begin{eqnarray}
\rho _{vac}(r_{p})=\rho _{vac}(t_{p})=\rho _{vac,0}\frac{R_{H,0}^{2}}{%
r_{p}^{2}}= \cr
\rho _{vac,0}\frac{ct_{0}^{2}}{ct_{p}^{2}}= \rho _{vac,0}\frac{%
t_{0}^{2}}{t_{p}^{2}}.  \label{35}
\end{eqnarray}

If we now substitute $\rho_{vac,0}$ by $3/8\pi G t_0^2$ (ref. Eq. (\ref{30}%
)) then Eq. (\ref{35}) can be written as:

\begin{equation}
\rho_{vac}(r_{p})=\rho_{vac}(t_{p})=\frac{3}{8\pi G t_0^2} \frac{t_0^2}{t_{p}^2} = 
\frac{3}{8\pi G t_{p}^2}.  \label{36}
\end{equation}

When we replace the Planck time $t_{p}=r_{p}/c=\sqrt{\hbar G/c^5}$ we finally
get the following formula:

\begin{equation}
\rho_{vac}(r_{p})=\rho_{vac}(t_{p})=\frac{3}{8\pi} \frac{c^5}{\hbar G^2} = \rho_{p}
\label{37}
\end{equation}

which is identical to the Planck density $\rho_{p}$ defined by the ratio of a
half Planck mass $\frac{1}{2}m_{p}=\frac{1}{2} \sqrt{\hbar c/G}$ and the
Planck volume $\frac{4\pi}{3} r_{p}^3$ with the Planck length $r_{p}=\sqrt{\hbar
G/c^3}$. This means that the equivalent vacuum mass density which scales
according to $R^{-2}$ in our model can be described as a scaling Planck
density $\rho_{p}$. In fact, we can re-write Eq. (\ref{40}) by replacing the factor $3/8 \pi G$ using Eq. (\ref{37}) and get:

\begin{equation}
\rho_{vac}(t)=\frac{3}{8 \pi G t^2}=\rho_{{p}}\frac{\hbar G/c^5}{t^2}=\rho_{p}\frac{t_{p}^2}{t^2}  \label{38}
\end{equation}

where the Planck time $t_{{p}}=\sqrt{\hbar G/c^5}$ is now the reference time.  The ratio $\rho_{vac,0}/\rho_{p}$ is
then simply given by:

\begin{equation}
\frac{\rho_{vac,0}}{\rho_{p}}=\frac{t_0^2}{t_{p}^2} \approx \cdot10^{-122}
\label{41}
\end{equation}

and also is a well-known discrepancy factor with respect to the ratio of the
present vacuum mass density on one hand and the theoretical value of the
vacuum mass density that follows from field-theoretical calculations on the
other hand (Zeldovich 1968; Weinberg 1989). Thus we can conclude, that this discrepancy vanishes
for a vacuum energy density that scales according to $R^{-2}$ as shown in
this paper.

\end{document}